# Scaling of population resilience with dispersal length and habitat size

Rodrigo Crespo-Miguel[a], Javier Jarillo[b], Francisco J. Cao-García[a,c,] [*]

[a] Departamento de Estructura de la Materia, Física Térmica y Electrónica, Facultad de Ciencias Físicas, Universidad Complutense de Madrid. Plaza de Ciencias 1, 28040 Madrid, Spain.

[b] Research Unit of Environmental and Evolutionary Biology, Namur Institute of Complex Systems, and Institute of Life, Earth, and the Environment, University of Namur, Rue de Bruxelles 61, Namur, 5000, Belgium.

[c] Instituto Madrileño de Estudios Avanzados en Nanociencia (IMDEA-Nanociencia). Calle Faraday 9, 28049 Madrid, Spain.

* Email: francao@ucm.es

## Abstract

Environmental fluctuations can create population-depleted areas and even extinct areas for the population. This effect is more severe in the presence of the Allee effect (decreasing growth rate at low population densities). Dispersal inside the habitat provides a rescue effect on population-depleted areas, enhancing the population resilience to environmental fluctuations. Habitat reduction decreases the effectiveness of the dispersal rescue mechanism. We report here how the population resilience to environmental fluctuations decreases when the dispersal length or the habitat size are reduced. The resilience reduction is characterized by a decrease of the extinction threshold for environmental fluctuations. The extinction threshold is shown to scale with the ratio between the dispersal length and the scale of environmental synchrony, i.e., it is the dispersal connection between non-environmentally-correlated regions that provides resilience to environmental fluctuations. Habitat reduction also decreases the resilience to environmental fluctuations, when the habitat size is similar to or smaller than the characteristic dispersal distances. The power laws of these scaling behaviors are characterized here. Alternative scaling functions with spatial scales of population synchrony are found to fit the simulations worse. These results support the dispersal length as the critical scale for extinction induced by habitat reduction.

## 1. Introduction

Habitat reduction is one of the main causes of danger for population stability (Fahrig 1997). The extinction risk is higher for species that experience (strong) Allee Effect (decreasing growth rate at low population densities) (Allee and Schuett 1927; Allee 1931) because habitat reduction emphasizes the harmful phenomena for small population densities. Genetic variability is reduced due to the increasing inbreeding in small habitats (Wagenius, Lonsdorf, and Neuhauser 2007; Bruggeman, Wiegand, and Fernández 2010). The weakest individuals of strongly territorial or intra-competing species can be displaced and die because they cannot find a place to settle down without being attacked by their congeners (Jager, Carr, and Efroymson 2006). Other possible consequences of habitat reduction include species that suffer from lack of food and are forced to change their diet (Araújo et al. 2014). Several studies also prove the detrimental effect of fragmentation on populations. For example, stochastic simulations in a bidimensional space in which each individual moves, dies, or procreates randomly every time step (Fahrig 1997) show that fragmentation and even more habitat loss imply a greater extinction risk. Stochastic logistic growth models have also proven that fragmentation reduces population abundance (Herbener, Tavener, and Hobbs 2012), i.e., the sum of the population size in two separate patches is always lower than the population when the patches are together. Here, we will mainly explore the impact of habitat size reduction by decreasing the effective dispersal length.

Environmental fluctuations represent stochastic external factors (as weather fluctuations) influencing the population dynamics. Environmental fluctuations typically have spatial synchrony and lead to spatial synchrony on the population fluctuations. These spatially synchronized population fluctuations imply simultaneous population depletions or even local or global extinctions (Heino et al. 1997; Engen, Lande, and Sæther 2002; Engen 2007). The spatial scale of population synchrony is equal (Moran 1953) or larger, due to dispersal (Lande, Engen, and Sæther 1999), than the spatial scale of synchrony of environmental fluctuations. Interspecies interactions can further increase the spatial scale of population synchrony (Bjørnstad, Ims, and Lambin 1999; Blasius, Huppert, and Stone 1999; Cazelles and Boudjema 2001; Ripa and Ranta 2007; Jarillo et al. 2018, 2020; Fernández-Grande and Cao-García 2020). Here, we describe the risk of extinction in terms of population resilience to environmental fluctuations. We define the extinction threshold (for the environmental fluctuations) as the value of the amplitude of environmental fluctuations above which environmental fluctuations cause a global extinction of the population (Crespo and Cao-García 2020). The extinction threshold provides a measure of the resilience to extinction.

In spatially extended populations (also known as metapopulations), dispersal provides a rescue mechanism to prevent local extinction from becoming global. Dispersal allows repopulating depleted regions with individuals from the non-depleted areas. Theoretical studies have shown that dispersal is much more effective as a mechanism to enhance resilience to environmental fluctuation if the area occupied by the population is much larger than the scale of population synchrony (Engen, Lande, and Sæther 2002). In addition, habitat fragmentation or habitat reduction effectively decreases dispersal as it has been observed in field studies for animals such as squirrels (Antolin et al. 2001) and in simulations that mimic the natural growth of populations of different species of plants (Collingham and Huntley 2000; Dullinger et al. 2015).

Here, we study the impact on resilience to environmental fluctuations due to reducing the dispersal length. We also investigate the harmful consequences of habitat reduction as an effective limiter of dispersal length. See Section 3. The study is performed with the spatially extended population model (with Allee effect, dispersal, and stochastic environmental fluctuations) introduced in Section 2.

# 2. Methods: Spatially extended population model for finite and infinite habitats

To study the effects of habitat size in population dynamics, we introduce a spatially extended population model with Allee effect, dispersal, and environmental stochasticity. Dispersal provides the population with resilience to environmental stochasticity for infinite habitats (Crespo and Cao-García 2020). Here, we check how this resilience is limited when the population is confined to a finite habitat, implemented through a finite simulation box with reflecting boundary conditions.

## 2.1. Infinite habitat

The evolution equation has both a deterministic and a stochastic part (caused by environmental stochasticity), giving the dynamics of the population density $N(x,t)$ as a function of the spatial location $x$ for future times $t$. The local deterministic dynamics is described by an Allee growth equation (Allee and Rosenthal 1949). Additionally, environmental stochasticity is added as a multiplicative noise, so the local dynamics of the population is described by

$$dN|_{local} = rN\left(\frac{N}{A} - 1\right)\left(1 - \frac{N}{K}\right) dt + \sigma N dB . \qquad (1)$$

Here, $r$ is the population's characteristic extinction rate (at low populations) and $K$ is the population's carrying capacity (the stable, viable local population density). $A$ is the population's Allee threshold, the minimum viable population density (i.e., the minimum local population density that gives a deterministic positive growth). The environmental stochastic term $\sigma N dB$ is proportional to the population density $N$. The amplitude of environmental fluctuations is given by $\sigma$, and $dB$ is a normalized Gaussian random field with zero mean $\langle dB(x,t)\rangle = 0$, and a spatial scale of synchrony of the fluctuations equal to $l_e$, which means that the environmental fluctuations are correlated within a length $l_e$ (Lande et al. 2003). This Gaussian field is uncorrelated in time and has an exponentially decreasing spatial correlation function

$$c_{dBdB}(y) = e^{-\sqrt{2}|y|/l_e}. \qquad (2)$$

This means $\langle dB(x,t)dB(x+y,t+\tau)\rangle = c_{dBdB}(y)\,\delta(\tau)\,dt$, with $\delta(\tau)$ the Dirac delta.

Dispersal is also considered, as it plays a crucial role in the persistence of a spatially extended population (Crespo and Cao-García 2020). If we consider individuals dispersing with a dispersal rate $m$ to a mean characteristic distance equal to $l_m$, then the dispersal term can be described by

$$dN|_{dispersal} = -m\,N\,dt + m\,dt \int N(y,t)f(x-y)dy , \qquad (3)$$

Where the function $f(x-y)$ is a Gaussian function with mean zero and variance equal to $l_m^2$. This equation describes that individuals leave from a particular point $y$ with probability $m\,dt$, and they disperse with probability $f(x-y)$ to a distance $x-y$, usually of the order of $l_m$, coupling the population densities over space.

Combining Eq. (1) and (3), we get the dynamical equation of a spatially extended, non-confined population,

$$dN = rN\left(\frac{N}{A} - 1\right)\left(1 - \frac{N}{K}\right) dt - m N \, dt + m \, dt \int N(y,t) f(x-y) dy + \sigma N dB \,. \qquad (4)$$

This non-confined population equation simulates a population on an infinite habitat, corresponding to an infinite confinement size, $L = \infty$. In practice, this corresponds to cases where the size of the habitat is much larger than the characteristic scales of the systems (in particular, larger than the population synchrony scales). Numerically, we considered large confinement size $L$, and periodic boundary conditions to reduce the border effects. (Our results revealed that border effects are negligible for values of $L \gg l_m$ as we discuss later in detail in the results section.)

The deterministic local term, the first term in Eq. (4), has two stable population densities, $N = 0$ (extinction) and $N = K$ (carrying capacity). For small population fluctuations, the rates of return to the extinction or the carrying capacity states are given by $\gamma_0 = r$ and $\gamma_K = r\left(\frac{K}{A} - 1\right)$, respectively (Crespo and Cao-García 2020). For populations close to a stable equilibrium point, small environmental fluctuations lead to a spatial scale of population synchrony $l$ greater than the spatial scale of synchrony of environmental fluctuations $l_e$, with $l = \sqrt{l_e^2 + m l_m^2/\gamma}$ and $\gamma$ the rate of return to the stable equilibrium (Lande, Engen, and Sæther 1999). Therefore, in the present case, we can define two characteristic scales of population synchrony around extinction and around carrying capacity, $l_0$ and $l_K$, respectively, defined as

$$l_0 = \sqrt{l_e^2 + \frac{m \cdot l_m^2}{\gamma_0}}, \qquad (5)$$

$$l_K = \sqrt{l_e^2 + \frac{m \cdot l_m^2}{\gamma_K}}, \qquad (6)$$

corresponding to the different return rates to extinction and carrying capacity, $\gamma_0$ and $\gamma_K$, respectively (Crespo and Cao-García 2020).

## 2.2. Finite habitat

Here we aim to address the effects of habitat size on ecosystem resilience to environmental fluctuations. This provides very relevant information on the impact of habitat reduction or habitat fragmentation in ecosystems.

We consider a population confined between the frontier positions $x = a$ and $x = b$, i.e., in the interval $x \in [a, b]$, which means a confinement length of $L = b - a$. We consider reflecting boundary conditions in these two frontier positions. One of the possible ways to introduce these reflecting boundary conditions is to generalize the convolution $\int N(y,t) f(x-y) dy$ in Eq. (4) to a convolution $F_\infty^{[a,b]}(x)$ defined by the following iterative process. We start computing

$$F_0^{[a,b]}(x) = \int_a^b N(y,t) f(x-y) dy \qquad (7)$$

in an interval wider than $[a, b]$ by several $l_m$, for example $[a - 3l_m, b + 3l_m]$. Then, we alternatively reflect the dispersal tails outside each of the sides of the interval to the interior. Alternatively, applying the right frontier reflection transformation

$$F_{2n+1}^{[a,b]}(x) = \begin{cases} F_{2n}^{[a,b]}(x) + F_{2n}^{[a,b]}(2b - x) & \text{if } x \leq b \\ 0 & \text{if } x > b \end{cases}, \quad (8)$$

and the left frontier reflection transformation

$$F_{2n+2}^{[a,b]}(x) = \begin{cases} F_{2n+1}^{[a,b]}(x) + F_{2n+1}^{[a,b]}(2a - x) & \text{if } x \geq a \\ 0 & \text{if } x < a \end{cases}. \quad (9)$$

The process finally converges to a convolution $F_{\infty}^{[a,b]}(x)$, which gives the dispersal of the population confined in the interval $[a, b]$ with reflecting boundary conditions. The dynamical equation for the population confined in this interval is given by

$$dN = rN\left(\frac{N}{A} - 1\right)\left(1 - \frac{N}{K}\right) dt - m\, N\, dt + m\, F_{\infty}^{[a,b]}\, dt + \sigma N dB\,. \quad (10)$$

Numerical simulations have been performed for finite habitat (finite $L$) using Eq. (10) with reflecting boundary conditions, and for infinite habitat ($L = \infty$) using Eq. (4) with periodic boundary conditions. Spatial and temporal resolutions are implemented in the same way for both cases. We consider the environmental synchrony scale, $l_e$, as the reference length, and set the spatial resolutions to have at least 20 lattice nodes per minimum parameter length (i.e., the smallest parameter among $l_e$, $l_m$, and $L$). The time resolution has been set as 50 times smaller than the minimum of the two characteristic time scales, $r^{-1}$ and $m^{-1}$ (the inverses of the extinction rate $r$ and the dispersal rate $m$ respectively). The extinction threshold for the amplitude of environmental fluctuations is obtained from several simulations stopping at 100 times $r^{-1}$ (which is also a large time compared to the characteristic dispersal time, as for the simulations performed $r^{-1} \sim 10\, m^{-1}$). (The extinction threshold mildly depends on the logarithm of the maximum simulation time as discussed in Appendix C. Increasing the maximum simulation time one order of magnitude, from $t = 100r^{-1}$ to $t = 1000r^{-1}$, only implies a change of the order of 10% in the extinction threshold.) Table I summarizes the variables used in this article, their definitions and units.

| Variables | Description |
|---|---|
| $N(x, t)$ | Population density at a given position x and time t. Units of space$^{-1}$ |
| $A$ | Allee threshold of the species. Species with a population density lower than $A$ has negative growth in a deterministic system. Units of space$^{-1}$ |
| $K$ | Carrying capacity of the species. Units of space$^{-1}$ |
| $r$ | Extinction rate of the species (at low population). Units of time$^{-1}$ |
| $\gamma_0$ | Rate of return to extinction, $\gamma_0 = r$. Units of time$^{-1}$ |
| $\gamma_K$ | Rate of return to carrying capacity, $\gamma_K = r\,(K/A - 1)$. Units of time$^{-1}$ |
| $m$ | Dispersal rate of the species. Units of time$^{-1}$ |
| $l_e$ | Spatial scale of synchrony of environmental fluctuations. Units of length |
| $l_m$ | Characteristic dispersal distances of the population. Units of length |
| $L$ | Confinement size. Units of length |
| $\sigma$ | Amplitude of the environmental fluctuations. It is equal to the standard deviation of the environmental fluctuations. Units of time$^{-1/2}$ |
| $\sigma_{extinction}$ | Extinction threshold for the amplitude of environmental fluctuations (Minimum amplitude of the environmental fluctuations that ensures global extinction). Units of time$^{-1/2}$ |

*Table 1: Variables used in this article, definitions, and units.*

## 3. Resilience to environmental fluctuations

Environmental fluctuations can lead an otherwise stable population to global extinction. Here, we address how large are the environmental fluctuations that a population can endure without going extinct. We define the extinction threshold $\sigma_{extinction}$ as the minimum amplitude of environmental fluctuations that ensures global extinction (See Appendix D). The extinction threshold provides a measure of the resilience of the population to environmental fluctuations.

Dispersal plays an essential role in making the populations more resilient to environmental fluctuations, recovering depleted or extinct populations in one location through the dispersal of the individuals from nearby non-depleted areas (Hanski and Gyllenberg 1993; Gotelli 1991; Crespo and Cao-García 2020). Habitat reduction (or habitat fragmentation) confines the population to a smaller region, where, therefore, the population fluctuations are more correlated. This higher population correlation reduces the effectiveness of this dispersal recovery mechanism. As a result, habitat reduction decreases the resilience of the population to environmental fluctuations. See Figs. 1 and 2. Reductions in the dispersal length $l_m$ also lead to reductions in the resilience of the population to environmental fluctuations, as we see in Fig. 2, and we study below in more detail in Section 3.1, where we show the scaling of the extinction threshold $\sigma_{extinction}$ with the dispersal length $l_m$. In Section 3.2, we characterize the scaling of the extinction threshold $\sigma_{extinction}$ with habitat size $L$. Finally, in Section 3.3, we show how we can define an effective dispersal length $l_{m,effective}(L)$ to describe how the dispersal length $l_m$ is effectively reduced as the habitat size $L$ decreases. This effective reduction in $l_m$ leads to earlier saturation of the extinction threshold as a function of $l_m$ for smaller confinement sizes $L$, as Fig. 2 shows

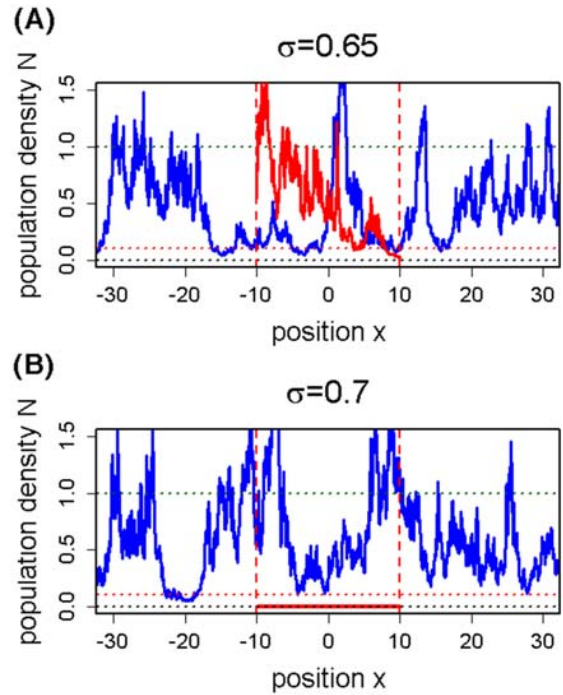

*Figure 1: Habitat reduction can lead to extinction, as it reduces the resilience to environmental fluctuations. Population density $N(x)$ at long times ($t = 100\ r^{-1}$) for a huge confinement size, $L = 142$ (blue solid line) and for a medium confinement size, $L = 20$ (red solid line), and for two amplitudes of the environmental noise $\sigma = 0.65$ (upper panel) and $\sigma = 0.70$ (bottom panel). The figure shows an example of how the reduction of confinement size reduces the resilience to environmental fluctuations, leading to the extinction of the more confined population when environmental noise is increased (red solid line of the bottom panel). Horizontal dotted lines represent extinction $N = 0$ (black), Allee threshold $N = A = 0.1$ (red), and carrying capacity $N = K = 1$ (green) population density values. These parameter values are common to all the cases represented in this figure, and also the extinction rate $r = 0.1$, the dispersal rate $m = 1$, and the dispersal length $l_m = l_e$ (with the spatial scale of environmental synchrony $l_e$ chosen as the unit of length $l_e = 1$). Huge confinement size $L = 142$ (blue solid line) was simulated with Eq. (3) and periodic boundary conditions, while medium confinement size $L = 20$ (red solid line) was simulated with Eq. (9) and reflecting boundary conditions.*

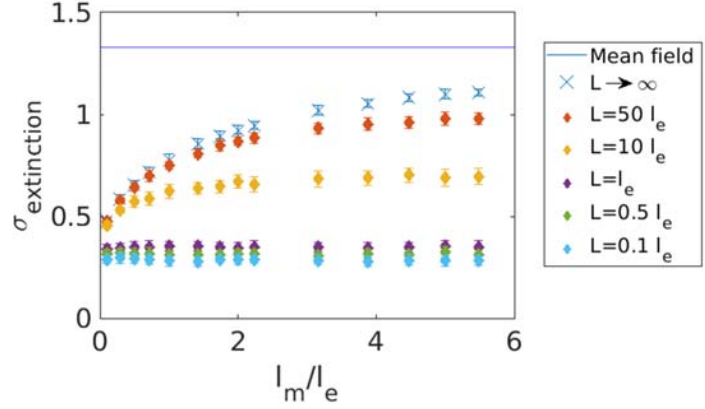

***Figure 2: Extinction threshold $\sigma_{extinction}$ as a function of the ratio $l_m/l_e$ for different confinement sizes L.*** *X-shaped points indicate simulations with a huge confinement size $L = 100\sqrt{l_e^2 + l_m^2}$ and periodic boundary conditions, whereas diamond-shaped points indicate simulations with a finite confinement size and reflecting boundary conditions. Vertical bars indicate uncertainty in the simulation result (see Appendix D). All simulations are for Allee threshold $A = 0.1$, carrying capacity $K = 1$, extinction rate $r = 0.1$, dispersal rate $m = 1$, and spatial scale of synchrony of environmental fluctuations $l_e = 1$. The blue solid line represents the result in the limit of infinite habitat size and characteristic dispersal length, $L = \infty, \frac{l_m}{l_e} = \infty$ and it has been obtained using the mean-field approximation as described in Ref. (Crespo and Cao-García 2020).*

### 3.1. Greater dispersal lengths increase the resilience to environmental fluctuations

Increasing the dispersal length $l_m$ favors the dispersal rescue effect for the population. This enhanced rescue effect enlarges the extinction threshold $\sigma_{extinction}$ (see Fig. 2) with $l_m$, until it saturates to the mean-field value ($l_m \to \infty$). Fig. 2 also shows that the extinction threshold depends additionally on the habitat size $L$, i.e., $\sigma_{extinction}(l_m, L)$. However, we will focus first on the dependence on the dispersal length $l_m$ for infinite habitats ($L = \infty$). Therefore, we will study the extinction threshold for infinite habitats $\sigma_{extinction,\infty}(l_m) = \sigma_{extinction}(l_m, L \to \infty)$. The mean-field value for an infinite habitat is denoted here by $\sigma_\infty^\infty = \sigma_{extinction}(l_m \to \infty, L \to \infty)$ and it is computed by the mean-field approximation explained in Ref. (Crespo and Cao-García 2020). The ration of extinction threshold for infinite habitat $\sigma_{extinction,\infty}$ and its respective mean-field value $\sigma_\infty^\infty$ for each dispersal rate $m$ is plotted as a function of the ratio of the dispersal length and the environmental correlation length, $l_m/l_e$, in Fig. 3. This two ratios show an approximate scaling of the form

$$\frac{\sigma_{extinction,\infty}\left(m, \frac{l_m}{l_e}\right)}{\sigma_\infty^\infty(m)} = M\left(\frac{l_m}{l_e}\right) = \left(\frac{1}{1 + \left(b_M \frac{l_e}{l_m}\right)^{n_M}}\right)^{1/d_M}. \qquad (11)$$

A maximum likelihood fit (see Appendix A) of this scaling form to the simulation results leads to the values of the fitting parameters: $d_M$, $n_M$, and $b_M$ indicated in Table 2. Plots of $\sigma_{extinction,\infty}/\sigma_\infty^\infty$ as a function of the ratios $l_m/l_0$ or $l_m/l_K$ do not present this approximate scaling behavior. See Fig. B1 in Appendix B.

| $n_M$ | $d_M$ | $b_M$ | SE | $-\ln \mathcal{L}$ | $AIC_C$ |
|---|---|---|---|---|---|
| 1.9±1.4 | 8.0±6.8 | 10.1±3.2 | 0.105 | 146 | 298 |

*Table 2: Parameters giving the maximum likelihood fit for the 42 points in Fig. 3 to Eq. (11). Uncertainties have been calculated with a confidence interval of 68%, i.e., at one-sigma level. Squared error SE, logarithm of the likelihood ($-\ln \mathcal{L}$) and Akaike Information Criterion $AIC_C$ are also included (See Appendix A for their definitions).*

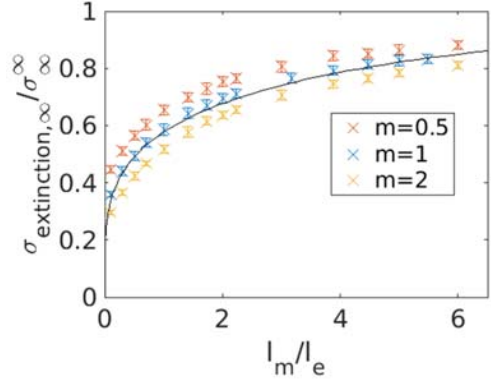

*Figure 3: Extinction threshold for infinite habitat $\sigma_{extinction,\infty}$ as a function of the dispersal length $l_m$, for various values of the dispersal rate m.* The figure shows that the extinction threshold for infinite habitat $\sigma_{extinction,\infty}$ (divided by the respective mean-field value $\sigma_\infty^\infty$ for each dispersal rate $m$) is fitted by the scaling function in Eq. (11) with the parameter values in Table 2 (solid black line). The dispersal length is expressed in units of the spatial scale of environmental fluctuations $l_e$. Vertical bars indicate uncertainty in the simulation results (see Appendix D). All simulations are for Allee threshold $A = 0.1$, carrying capacity $K = 1$, extinction rate $r = 0.1$, and spatial scale of synchrony of environmental fluctuations $l_e = 1$.

The data in Fig.3 have a slight dependence in the dispersal rate $m$, showing that points with smaller dispersal rates are higher than those with larger dispersal rate for the same ratio $l_m/l_e$. Thus, the extinction threshold $\sigma_{extinction}$ divided by its respective mean-field value $\sigma_\infty^\infty$, has a better fit with a function scaling with $\left(\frac{m}{r}\right)^{s_M}\left(\frac{l_m}{l_e}\right)$, where $s_M$ is an additional fitting parameter,

$$\frac{\sigma_{extinction,\infty}\left(m, \frac{l_m}{l_e}\right)}{\sigma_\infty^\infty(m)} = M\left(\frac{m}{r}, \frac{l_m}{l_e}\right) = \left(\frac{1}{1 + \left(b_M \frac{l_e}{l_m}\left(\frac{r}{m}\right)^{s_M}\right)^{n_M}}\right)^{1/d_M}. \quad (12)$$

The fit to the numerical results of this scaling law, Eq. (12), is better than for the previous one, Eq. (11), even taking into account the AIC penalty for the additional parameter ($s_M$). See Fig. 4 and Table 3. The significant difference in $AIC_C$ yielded by fitting Eqs. (11) and (12), $\Delta AIC_C =$ 481 ≫ 10, implies that the model with higher AIC$_C$, given by Eq. (11), can be discarded entirely (Appendix A). The best-fitting model (the model with the lowest AIC$_C$), Eq. (12), indicates that the mean-field value is reached for long dispersal distances $l_m$, but more slowly for higher dispersal rates $m$.

| n$_M$ | d$_M$ | b$_M$ | s$_M$ | SE | $-\ln \mathcal{L}$ | AIC$_C$ |
|---|---|---|---|---|---|---|
| 1.26±0.21 | 5.0±1.1 | 1.57±0.36 | -0.667±0.049 | 0.0120 | -96.0 | -183 |

**Table 3: Parameters giving the maximum likelihood fit for the 42 points in Fig. 4 to Eq. (12).** Uncertainties have been calculated with a confidence interval of 68%, i.e., at one-sigma level. Squared error (SE), logarithm of the likelihood ($-\ln \mathcal{L}$) and Akaike Information Criterion (AIC$_C$) are also included (See Appendix A for their definitions).

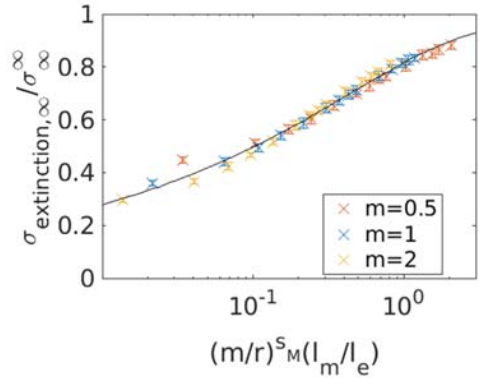

*Figure 4: Extinction threshold for infinite habitat $\sigma_{extinction,\infty}$ scales as a function of the ratio $\left(\frac{m}{r}\right)^{s_M}\left(\frac{l_m}{l_e}\right)$, with $s_M = -0.667$. The figure shows that the extinction threshold for infinite habitat $\sigma_{extinction,\infty}$ (divided by the respective mean-field value $\sigma_\infty^\infty$ for each dispersal rate $m$) is fitted by the scaling function in Eq. (12) with the parameter values in Table 3 (solid black line). Vertical bars indicate uncertainty in the simulation results (see Appendix D). All simulations are for Allee threshold $A = 0.1$, carrying capacity $K = 1$, extinction rate $r = 0.1$, and spatial scale of synchrony of environmental fluctuations $l_e = 1$.*

### 3.2. Habitat reduction decreases resilience to environmental fluctuations

As we explained above, the resilience to environmental fluctuations can be characterized by the minimum amplitude of environmental fluctuations that leads to global extinction, the extinction threshold $\sigma_{extinction}(m, l_m, L)$. The extinction threshold decreases as the habitat size decreases (for example, due to habitat fragmentation). We characterized in the previous section the dependence of the extinction threshold with the dispersal length $l_m$ for infinite habitats. Using these previous results and an analogous approach, we will find the scaling behavior with the habitat size $L$. We consider the ratio of the extinction threshold for finite and infinite habitat size (see Fig. 5) and compare fits of the type

$$\frac{\sigma_{extinction}(m, l_m, L)}{\sigma_{extinction,\infty}\left(m, \frac{l_m}{l_e}\right)} = F\left(\frac{L}{l_i}\right) = \left(\frac{1}{1 + \left(b_F \frac{l_i}{L}\right)^{n_F}}\right)^{1/d_F}, \qquad (13)$$

where $l_i$ can represent the spatial scale of environmental synchrony $l_e$, the dispersal length $l_m$, the spatial scale of population synchrony close to extinction $l_0 = \sqrt{l_e^2 + ml_m^2/\gamma_0}$, or the spatial scale of population synchrony close to the carrying capacity $l_K = \sqrt{l_e^2 + ml_m^2/\gamma_K}$ [as introduced in Eqs. (5) and (6) above]. The maximum likelihood fitting of Eq. (12) to the simulation results (Fig. 5) gives the parameter values indicated in Table 3, where the AIC$_C$ results indicate that the best fitting model is that with $l_i = l_m$, closely followed by $l_i = l_0$. The other models can be discarded as its difference in AIC$_C$ is greater than 10 (Burnham and Anderson 2002). In particular, the model with $l_i = l_e$ is completely discarded due to the large value of AICc, and the large uncertainty and instability in the parameter value determination, which is consistent with the dispersal of the values observed in Fig. 5 for this case.

|        | $n_F$     | $d_F$    | $b_F$     | SE    | $-\ln \mathcal{L}$ | AIC$_C$ |
|--------|-----------|----------|-----------|-------|---------|---------|
| L/l$_m$ | 1.93±0.83 | 8.5±4.0  | 28.7±6.4  | 0.158 | -69.4   | -132    |
| L/l$_0$ | 1.84±0.73 | 8.0±3.5  | 6.4±1.5   | 0.193 | -68.9   | -131    |
| L/l$_K$ | 2.3±1.6   | 10.1±7.6 | 16.5±4.6  | 0.382 | -21.6   | -36.8   |
| L/l$_e$ | 3.0±4.0   | 14±19    | 58±26     | 1.033 | 209     | 424     |

*Table 4: Parameters giving the maximum likelihood fit for the 60 points of Fig. 5 to Eq. (13). Uncertainties have been calculated with a confidence interval of 68%, i.e., at one-sigma level. Squared error SE, logarithm of the likelihood ($-\ln \mathcal{L}$) and Akaike Information Criterion AIC$_C$ are also included (See Appendix A for their definitions).*

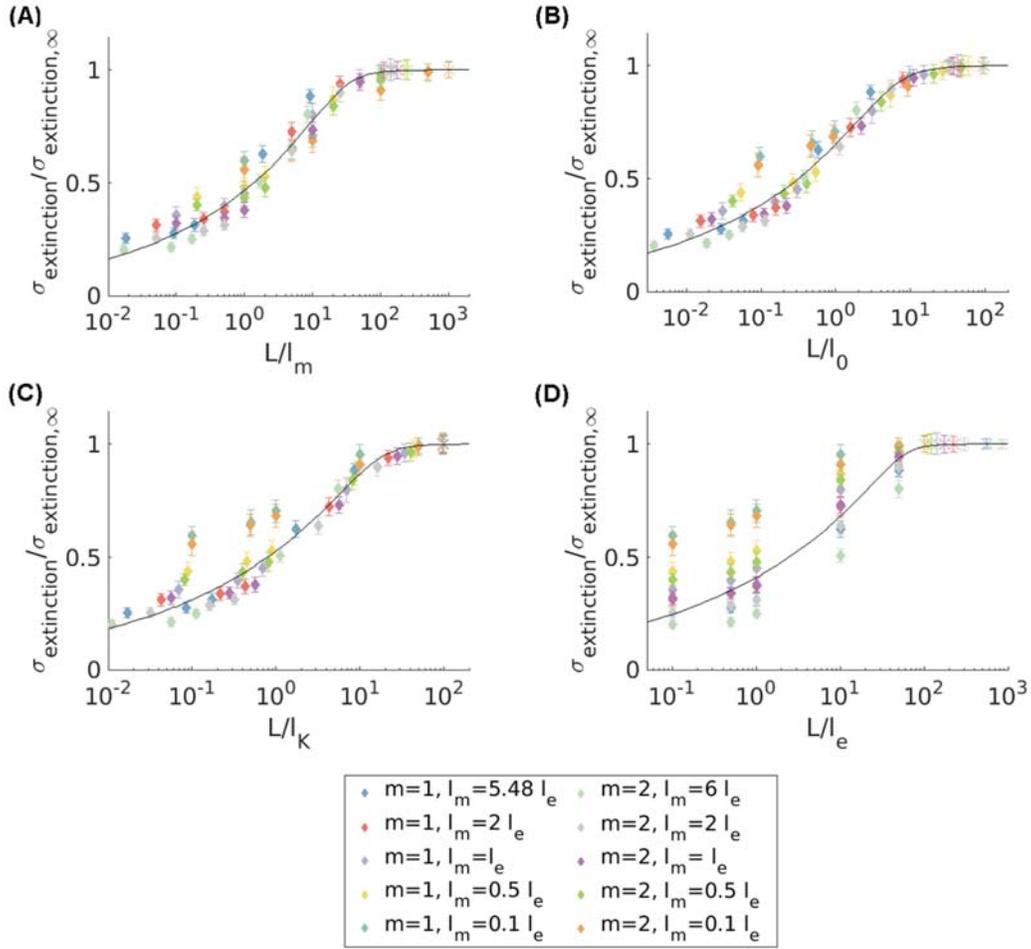

*Figure 5: Extinction thresholds for fragmented habitats $\sigma_{extinction}$ as functions of the size of the habitat L divided by the dispersal length of the population $l_m$ (Panel A), the spatial scale of population synchrony around extinction $l_0$ (Panel B), the spatial scale of population synchrony around carrying capacity $l_K$ (Panel C), and the spatial scale of synchrony of environmental fluctuations $l_e$ (Panel D). The figure shows that the extinction threshold for a fragmented habitat $\sigma_{extinction}$ divided by the respective infinite habitat value $\sigma_{extinction,\infty}$ (for the same dispersal rate m and dispersal length $l_m$) is fitted by the scaling function in Eq. (13) with the parameter values in Table 4 (solid black line). Vertical bars indicate uncertainty in the numerical results (see Appendix D). All simulations are for Allee threshold A =0.1, carrying capacity K =1, extinction rate r =0.1, and spatial scale of synchrony of environmental fluctuations $l_e$ =1.*

Additionally, the improved fit found in the previous subsection for the $l_m/l_e$ scaling introducing a dependence on the ratio between the dispersal rate $m$ and the extinction rate $r$ motivates us to perform a similar check for the scaling of the habitat size $L$. We found that the data in Fig. 5 scale better as a function of $\left(\frac{m}{r}\right)^{S_F}\left(\frac{L}{l_i}\right)$,

$$\frac{\sigma_{extinction}(m, l_m, L)}{\sigma_{extinction,\infty}\left(m, \frac{l_m}{l_e}\right)} = F\left(\frac{m}{r}, \frac{L}{l_i}\right) = \left(\frac{1}{1 + \left(b_F \frac{l_i}{L}\left(\frac{r}{m}\right)^{s_F}\right)^{n_F}}\right)^{1/d_F}, \quad (14)$$

which fits the numerical results with the parameters of Table 5 (Fig. 6). We see, comparing Table 4 and 5, that the AIC$_C$ is much lower for a maximum likelihood fit to Eq. (14) than to Eq.(13) ($\Delta AIC_C > 10$), therefore the model with more empirical support is that which follows Eq. (14) with $l_i = l_m$. The scaling given by Eq. (14) indicates that the infinite habitat value is reached for long ratios $L/l_i$, and that limit is reached slower for higher dispersal rates $m$.

|  | n$_F$ | d$_F$ | b$_F$ | s$_F$ | SE | $-\ln \mathcal{L}$ | AIC$_C$ |
|---|---|---|---|---|---|---|---|
| L/l$_m$ | 1.91±0.70 | 8.5±3.4 | 2.5±1.4 | -0.93±0.21 | 0.132 | -89.6 | -170 |
| L/l$_0$ | 1.87±0.72 | 8.2±3.5 | 1.6 ±1.0 | -0.53±0.24 | 0.188 | -75.5 | -142 |
| L/l$_K$ | 2.2±1.4 | 9.7±6.6 | 3.1±2.6 | -0.62±0.29 | 0.367 | -31.1 | -53.5 |
| L/l$_e$ | 2.5±2.5 | 11±12 | 2.7±3.7 | -1.09±0.48 | 0.963 | 179 | 367 |

*Table 5: Parameters giving the maximum likelihood fit for the 60 points of Fig. 6 to the Eq. (14). Uncertainties have been calculated with a confidence interval of 68%, i.e., at one-sigma level. Squared error SE, logarithm of the likelihood ($-\ln \mathcal{L}$) and Akaike Information Criterion AIC$_C$ are also included (See Appendix A for their definitions).*

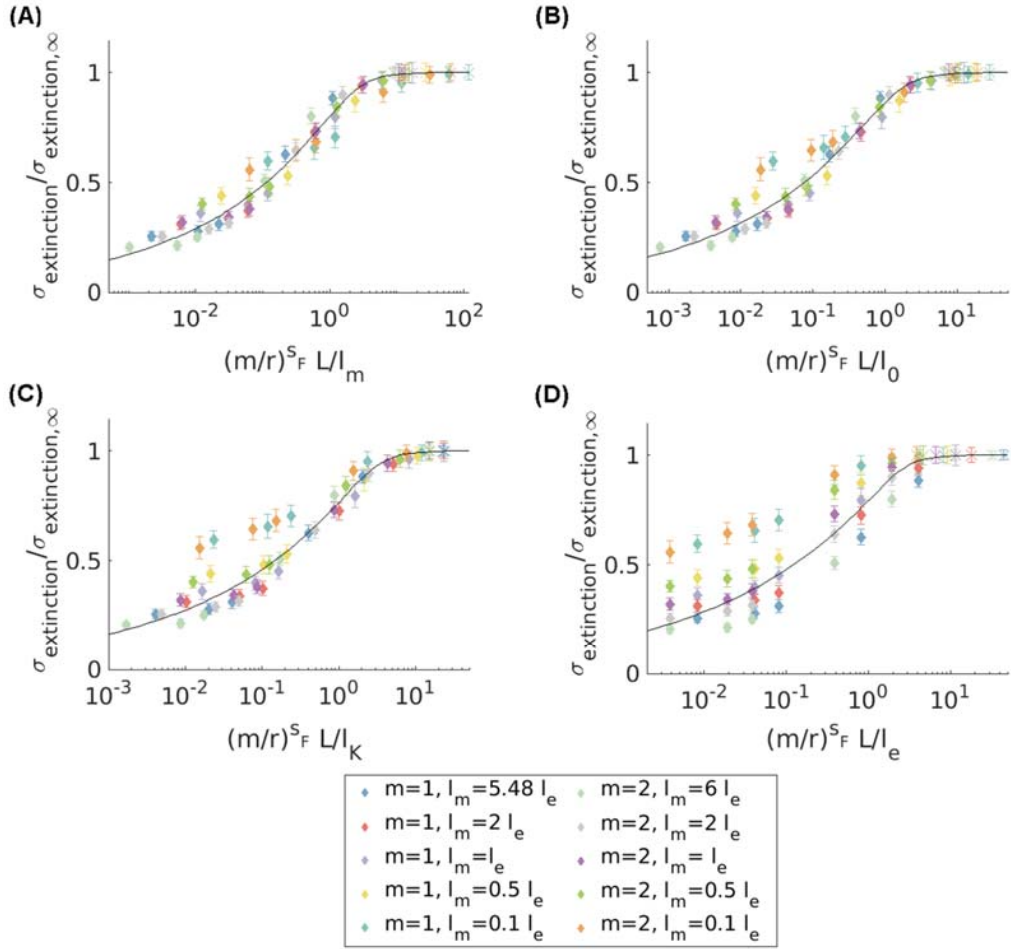

Figure 6: Extinction thresholds for fragmented habitats $\sigma_{extinction}$ as functions of the size of the habitat L multiplied by the ratio $\left(\frac{m}{r}\right)^{s_F}$ (value of $s_F$ is given in Table 5) and divided by the dispersal length of the population $l_m$ (Panel A), the spatial scale of population synchrony around extinction $l_0$ (Panel B), the spatial scale of population synchrony around carrying capacity $l_K$ (Panel C), and the spatial scale of synchrony of environmental fluctuations $l_e$ (Panel D). The figure shows that the extinction threshold for a fragmented habitat $\sigma_{extinction}$ (divided by the respective infinite habitat value $\sigma_{extinction,\infty}$ for the same dispersal rate m and dispersal length $l_m$) is fitted by an approximate scaling described by Eq. (14) with the parameter values in Table 5 (solid black line). Vertical bars indicate uncertainty in the simulation result (see Appendix D). All simulations are for Allee threshold A =0.1, carrying capacity K =1, extinction rate r =0.1, and spatial scale of synchrony of environmental fluctuations $l_e$ =1.

Putting together the results of this subsection and the previous one provides the complete scaling behavior of the extinction threshold $\sigma_{extinction}$

$$\sigma_{extinction}(m, l_m, L) = \sigma_\infty^\infty(m) \, M\left(\frac{m}{r}, \frac{l_m}{l_e}\right) F\left(\frac{m}{r}, \frac{L}{l_m}\right)$$

$$= \sigma_\infty^\infty(m) \left(\frac{1}{1 + \left(\frac{b_M \, l_e}{l_m}\left(\frac{r}{m}\right)^{s_M}\right)^{n_M}}\right)^{1/d_M} \left(\frac{1}{1 + \left(\frac{b_F \, l_m}{L}\left(\frac{r}{m}\right)^{s_F}\right)^{n_F}}\right)^{1/d_F}, \quad (15)$$

with the values of the parameters given in the first row of Tables 3 and 5.

## 3.3. Habitat reduction decreases the effective dispersal length

From the results in previous subsections, we got that habitat reduction effectively decreases the dispersal length, leading to a detriment in the population's resilience to environmental fluctuations. Thus, we can define an effective dispersal length $l_{m,eff}(m, l_m, L)$, as the dispersal length of a population in an infinite habitat ($L = \infty$) that has the same extinction threshold as a population with dispersal length $l_m$ confined in a fragmented habitat of size $L$, if all the other parameters affecting both populations are equal. This definition is equivalent to the following equation [expanded using Eq. (14)]:

$$\sigma_{extinction}(m, l_{m,eff}, L \to \infty) = \sigma_{extinction}(m, l_m, L)$$
$$\Rightarrow M\left(\frac{m}{r}, \frac{l_{m,eff}}{l_e}\right) F\left(\frac{L}{l_m} \to \infty\right) = M\left(\frac{m}{r}, \frac{l_m}{l_e}\right) F\left(\frac{m}{r}, \frac{L}{l_m}\right) \quad (16)$$
$$\Rightarrow M\left(\frac{m}{r}, \frac{l_{m,eff}}{l_e}\right) = M\left(\frac{m}{r}, \frac{l_m}{l_e}\right) F\left(\frac{m}{r}, \frac{L}{l_m}\right).$$

Thus, we got the following expression for the effective dispersal length $l_{m,eff}(m, l_m, L)$,

$$l_{m,eff}(m, l_m, L) =$$

$$b_M l_e \left(\frac{r}{m}\right)^{s_M} \left(\left(1 + \left(\frac{b_M \, l_e}{l_m}\left(\frac{r}{m}\right)^{s_M}\right)^{n_M}\right) \cdot \left(1 + \left(\frac{b_F \, l_m}{L}\left(\frac{r}{m}\right)^{s_F}\right)^{n_F}\right)^{\frac{d_M}{d_F}} - 1\right)^{-\frac{1}{n_M}}, \quad (17)$$

with the parameters given in the first row of Tables 3 and 5. The effective dispersal length, as shown in Fig. 7, is a monotonously increasing function with the habitat size $L$, and it tends to the real dispersal length $l_m$ when the habitat size becomes sufficiently large. Thus, habitat size reduction implies a decrease in the effective dispersal length.

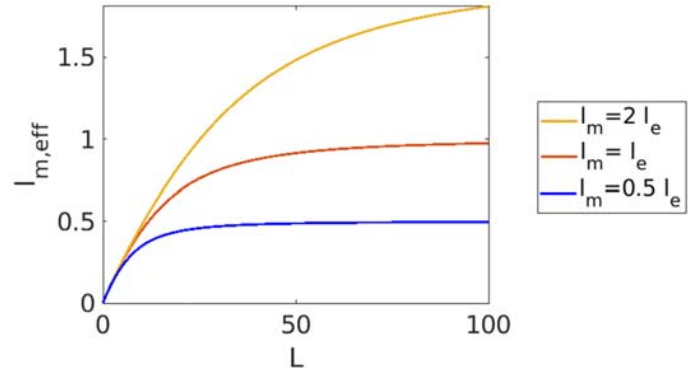

*Figure 7: Effective dispersal length $l_{m,eff}$, Eq. (17), as a function of the size of the habitat L for different dispersal distances $l_m = 0.5\, l_e$ (blue), $l_m = l_e$ (red) and $l_m = 2\, l_e$ (orange). All curves are for Allee threshold $A = 0.1$, carrying capacity $K = 1$, extinction rate $r = 0.1$, dispersal rate $m = 1$, and spatial scale of synchrony of environmental fluctuations $l_e = 1$.*

## 4. Discussion

We have shown that both habitat reduction and dispersal reduction decrease population resilience against environmental fluctuations. We have measured the population resilience with the extinction threshold $\sigma_{extinction}$, defined as the minimum amplitude of environmental fluctuations that ensures population extinction at long times (Appendix D). On the one hand, we have obtained that populations with dispersal distances larger than the scale of synchrony of environmental fluctuations are more resilient to fluctuations, and reach the largest extinction threshold for every confinement size. On the other hand, habitat reduction is found to decrease the resilience to environmental fluctuations, leading to a huge drop in the extinction threshold when the habitat size becomes of the order of the typical dispersal distances of the population.

These results imply that habitat reduction (or habitat fragmentation) causes an effective reduction in dispersal length, thus minimizing dispersal-induced resilience caused by the rescue effect. Dispersal is proved as an essential mechanism against extinction in populations with Allee Effect (Palmqvist and Lundberg 1998; Dennis et al. 2016; Crespo and Cao-García 2020). However, dispersal is not so effective when a smaller habitat size truncates its range. These results further clarify that habitat reduction (or habitat fragmentation) is a problem for endangered populations. Its impact can be much more significant for species that depend on diffusion when the habitat size is reduced close to its typical dispersal length. We can apply this knowledge by making a special effort to develop conservation strategies that include active prevention of habitat destruction, so environmental fluctuations are less likely to destroy endangered species populations. Deeper knowledge of populations' dynamics helps to optimize species conservation and sustainable exploitation policies.

Synchrony plays an essential role in the regional extinction risk (Heino et al. 1997; Engen, Lande, and Sæther 2002; Engen 2007). However, we found that the decrease of the extinction threshold with habitat size scales better with the ratio of the habitat size to the dispersal length than with the ratio of the habitat size to the spatial scale of synchrony. Thus, we think that the transition to regional extinction needs further study to achieve a complete understanding of the interplays among the spatial scale of population synchrony, the dispersal length, the habitat size, and the transitions to extinction.

Here, we have studied the one-dimensional case. The extension of this study to two dimensions would be interesting because many natural ecosystems are two-dimensional. It is known that the shapes of the fragmented patches can affect the survival of many species (Diamond 1975), favoring connectivity to support recovery by internal migration. Patch shapes can also determine colonized area expansion or contraction (Lewis and Kareiva 1993; Keitt, Lewis, and Holt 2001). Additionally, habitat reduction and fragmentation per se have slightly different effects on population resilience (Fahrig 2003, 1997). Studying the one- and two-dimensional patch structure and its connectivity can also provide insight into the transitions to extinction.

These results have been obtained for one-species systems, for which dispersal is beneficial, increasing the rescue effect. However, dispersal does not always benefit the dynamics for multiple-species systems since it can homogenize the habitats, removing some local species (Mouquet and Loreau 2003). Furthermore, the diversity is expected to peak at intermediate diffusion (Gravel, Massol, and Leibold 2016). Additionally, recent work has shown that coupling between heterogeneous patches can lead to a non-monotonic behavior of the extinction probability with the dispersal (Agranov and Bunin 2021). This work also stresses that dispersal from other patches might act as a non-gaussian noise in the receiving patch, extending the noise-induced phenomena found in ecosystems (Spagnolo, Valenti, and Fiasconaro 2004).

## Funding
This work was funded by European Regional Development Fund (ERDF) and the Spanish Ministry of Economy and Competitiveness through Grant RTI2018-095802-B-I00, and by European Union's Horizon 2020 through grant agreement No 817578 TRIATLAS. JJ was supported by CEFIC under the GETREAL project, and from the Fund for Scientific Research, FNRS (PDRT.0048.16).

## Author Contributions
RCM and FJCG conceived the manuscript. JJ wrote the original code used for the stochastic population dynamics. RCM adapted the code to populations with Allee effect and reflecting boundary conditions. RCM and JJ ran the numerical simulations. RCM and FJCG wrote the first version of the manuscript. RCM prepared the figures. All authors discussed the results and contributed to the final manuscript.

## References

Agranov, Tal, and Guy Bunin. 2021. "Extinctions of Coupled Populations, and Rare Event Dynamics under Non-Gaussian Noise." *Physical Review E* 104 (2): 1–14. https://doi.org/10.1103/PhysRevE.104.024106.

Akaike, Hirotugu. 1974. "A New Look at the Statistical Model Identification." *IEEE Transactions on Automatic Control* 19 (6): 716–23. https://doi.org/10.1109/TAC.1974.1100705.

Allee, W. C. 1931. *Animal Aggregations. A Study in General Sociology.* Chicago: The University of Chicago Press. https://doi.org/10.5962/bhl.title.7313.

Allee, W. C., and G. M. Rosenthal. 1949. "Group Survival Value for Philodina Rosola, A Rotifer." *Ecology* 30 (3): 395–97. https://doi.org/10.2307/1932623.

Allee, W. C., and J. F. Schuett. 1927. "Studies in Animal Aggregations: The Relation between Mass of Animals and Resistance to Colloidal Silver." *The Biological Bulletin* 53 (5): 301–17. https://doi.org/10.2307/1537055.

Antolin, Michael F., Beatrice Van Horne, Michael D. Berger, Jr., Alisha K. Holloway, Jennifer L. Roach, and Ronald D. Weeks, Jr. 2001. "Effective Population Size and Genetic Structure of a Piute Ground Squirrel (*Spermophilus Mollis*) Population." *Canadian Journal of Zoology* 79 (1): 26–34. https://doi.org/10.1139/cjz-79-1-26.

Araújo, Márcio S., R. Brian Langerhans, Sean T. Giery, and Craig A. Layman. 2014. "Ecosystem Fragmentation Drives Increased Diet Variation in an Endemic Livebearing Fish of the <scp>B</Scp> Ahamas." *Ecology and Evolution* 4 (16): 3298–3308. https://doi.org/10.1002/ece3.1140.

Bjørnstad, Ottar N., Rolf A. Ims, and Xavier Lambin. 1999. "Spatial Population Dynamics: Analyzing Patterns and Processes of Population Synchrony." *Trends in Ecology and Evolution* 14 (11): 427–32. https://doi.org/10.1016/S0169-5347(99)01677-8.

Blasius, Bernd, Amit Huppert, and Lewi Stone. 1999. "Complex Dynamics and Phase Synchronization in Spatially Extended Ecological Systems." *Nature* 399 (6734): 354–59. https://doi.org/10.1038/20676.

Bruggeman, Douglas J., Thorsten Wiegand, and Néstor Fernández. 2010. "The Relative Effects of Habitat Loss and Fragmentation on Population Genetic Variation in the Red-Cockaded Woodpecker (Picoides Borealis)." *Molecular Ecology* 19 (17): 3679–91. https://doi.org/10.1111/j.1365-294X.2010.04659.x.

Burnham, KP, and DR Anderson. 2002. "Model Selection and Multimodel Inference: A Practical Information-Theoretic Approach. 2nd Edn. Springer, Berlin." *Bayesian Data Analysis in Ecology Using Linear Models with R, BUGS, and STAN*.

Cazelles, Bernard, and Gérard Boudjema. 2001. "The Moran Effect and Phase Synchronization in Complex Spatial Community Dynamics." *The American Naturalist* 157 (6): 670–76. https://doi.org/10.1086/320624.

Collingham, Yvonne C., and Brian Huntley. 2000. "Impacts of Habitat Fragmentation and Patch Size upon Migration Rates." *Ecological Applications* 10 (1): 131–44. https://doi.org/10.1890/1051-0761(2000)010[0131:IOHFAP]2.0.CO;2.

Crespo, Rodrigo, and Francisco Javier Cao-García. 2020. "Dispersal-Induced Resilience to Stochastic Environmental Fluctuations in Populations with Allee Effect."

Dennis, Brian, Laila Assas, Saber Elaydi, Eddy Kwessi, and George Livadiotis. 2016. "Allee Effects and Resilience in Stochastic Populations." *Theoretical Ecology* 9 (3): 323–35. https://doi.org/10.1007/s12080-015-0288-2.

Diamond, Jared M. 1975. "The Island Dilemma: Lessons of Modern Biogeographic Studies for the Design of Natural Reserves." *Biological Conservation* 7 (2): 129–46. https://doi.org/10.1016/0006-3207(75)90052-X.

Dullinger, Stefan, Nicolas Dendoncker, Andreas Gattringer, Michael Leitner, Thomas Mang, Dietmar Moser, Caspar A. Mücher, et al. 2015. "Modelling the Effect of Habitat Fragmentation on Climate-Driven Migration of European Forest Understorey Plants." Edited by Ingolf Kühn. *Diversity and Distributions* 21 (12): 1375–87. https://doi.org/10.1111/ddi.12370.

Engen, Steinar. 2007. "Stochastic Growth and Extinction in a Spatial Geometric Brownian Population Model with Migration and Correlated Noise." *Mathematical Biosciences* 209

(1): 240–55. https://doi.org/10.1016/j.mbs.2006.08.011.

Engen, Steinar, Russell Lande, and Bernt-Erik Sæther. 2002. "The Spatial Scale of Population Fluctuations and Quasi-Extinction Risk." *The American Naturalist* 160 (4): 439–51. https://doi.org/10.1086/342072.

Fahrig, Lenore. 1997. "Relative Effects of Habitat Loss and Fragmentation on Population Extinction." *The Journal of Wildlife Management* 61 (3): 603. https://doi.org/10.2307/3802168.

———. 2003. "Effects of Habitat Fragmentation on Biodiversity." *Annual Review of Ecology, Evolution, and Systematics* 34 (1): 487–515. https://doi.org/10.1146/annurev.ecolsys.34.011802.132419.

Fernández-Grande, Miguel Ángel, and Francisco Javier Cao-García. 2020. "Spatial Scales of Population Synchrony Generally Increases as Fluctuations Propagate in a Two Species Ecosystem." *ArXiv*, 1–17.

Gotelli, Nicholas J. 1991. "Metapopulation Models: The Rescue Effect, the Propagule Rain, and the Core-Satellite Hypothesis." *The American Naturalist* 138 (3): 768–76. https://doi.org/10.1086/285249.

Gravel, Dominique, François Massol, and Mathew A. Leibold. 2016. "Stability and Complexity in Model Meta-Ecosystems." *Nature Communications* 7. https://doi.org/10.1038/ncomms12457.

Hanski, Ilkka, and Mats Gyllenberg. 1993. "Two General Metapopulation Models and the Core-Satellite Species Hypothesis." *The American Naturalist* 142 (1): 17–41. https://doi.org/10.1086/285527.

Heino, Mikko, Veijo Kaitala, Esa Ranta, and Jan Lindstrom. 1997. "Synchronous Dynamics and Rates of Extinction in Spatially Structured Populations." *Proceedings of the Royal Society B: Biological Sciences* 17 (1381): 481–86. https://doi.org/10.1098/rspb.1997.0069.

Herbener, Kathy W., Simon J. Tavener, and N. Thompson Hobbs. 2012. "The Distinct Effects of Habitat Fragmentation on Population Size." *Theoretical Ecology* 5 (1): 73–82. https://doi.org/10.1007/s12080-010-0097-6.

Hogg, David W, Jo Bovy, and Dustin Lang. 2010. "Data Analysis Recipes: Fitting a Model to Data," August.

Jager, Henriette I., Eric A. Carr, and Rebecca A. Efroymson. 2006. "Simulated Effects of Habitat Loss and Fragmentation on a Solitary Mustelid Predator." *Ecological Modelling* 191 (3–4): 416–30. https://doi.org/10.1016/j.ecolmodel.2005.05.025.

Jarillo, Javier, Bernt-Erik Erik Sæther, Steinar Engen, Francisco Javier Cao García, and Francisco Javier Cao-García. 2020. "Spatial Scales of Population Synchrony in Predator-Prey Systems." *American Naturalist* 195 (2): 216–30. https://doi.org/10.1086/706913.

Jarillo, Javier, Bernt Erik Sæther, Steinar Engen, and Francisco J. Cao. 2018. "Spatial Scales of Population Synchrony of Two Competing Species: Effects of Harvesting and Strength of Competition." *Oikos* 127 (10): 1459–70. https://doi.org/10.1111/oik.05069.

Keitt, Timothy H, Mark A Lewis, and Robert D Holt. 2001. "Allee Effects , Invasion Pinning , and Species ' Borders." *The American Naturalist* 157 (2): 203–16.

Lande, Russell, Steinar Engen, Bernt-Erik Sæther, and Bernt-Erik Saether. 2003. *Stochastic Population Dynamics in Ecology and Conservation*. *Oxford Series in Ecology and Evolution*.

Oxfo. Oxford: Oxford University Press. https://doi.org/10.1093/ISBN.

Lande, Russell, Steinar Engen, and Bernt-Erik Sæther. 1999. "Spatial Scale of Population Synchrony: Environmental Correlation versus Dispersal and Density Regulation." *The American Naturalist* 154 (3): 271–81. https://doi.org/10.1086/303240.

Lewis, M.A. A., and P. Kareiva. 1993. *Allee Dynamics and the Spread of Invading Organisms*. *Theoretical Population Biology*. Vol. 43. Academic Press. https://doi.org/10.1006/tpbi.1993.1007.

Moran, PAP Patrick Alfred Pierce. 1953. "The Statistical Analysis of the Canadian Lynx Cycle. II. Synchronization and Meteorology." *Australian Journal of Zoology* 1 (3): 291–98. https://doi.org/10.1071/ZO9530291.

Mouquet, Nicolas, and Michel Loreau. 2003. "Community Patterns in Source-Sink Metacommunities." *The American Naturalist* 162 (5): 544–57. https://doi.org/10.1086/378857.

Palmqvist, Eva, and Per Lundberg. 1998. "Population Extinctions in Correlated Environments." *Oikos* 83 (2): 359. https://doi.org/10.2307/3546850.

Ripa, Jörgen, and Esa Ranta. 2007. "Biological Filtering of Correlated Environments: Towards a Generalised Moran Theorem." *Oikos* 116 (5): 783–92.

Spagnolo, B., D. Valenti, and A. Fiasconaro. 2004. "Noise in Ecosystems: A Short Review." *Mathematical Biosciences and Engineering* 1 (1): 185–211. https://doi.org/10.3934/mbe.2004.1.185.

Wagenius, Stuart, Eric Lonsdorf, and Claudia Neuhauser. 2007. "Patch Aging and the S -Allee Effect: Breeding System Effects on the Demographic Response of Plants to Habitat Fragmentation." *The American Naturalist* 169 (3): 383–97. https://doi.org/10.1086/511313.

# Appendix A. Maximum likelihood fit and Akaike Information Criterion (AIC$_C$)

The fits used in the main text have been done by maximizing likelihood (Burnham and Anderson 2002). Likelihood is defined by equation A1 for a set of $n$ independent measures with equal standard deviation $S$,

$$-\ln \mathcal{L} = \frac{n}{2}\ln(2\pi) + \frac{n}{2}\ln S^2 + \frac{1}{2S^2}SE , \quad (A1)$$

where $SE = \sum_i^n (f(x_i) - y_i)^2$ is the total squared error, $f(x_i)$ is the value of the fit function for a parameter $x_i$, and $y_i$ its measured value.

This definition can be generalized for a set of measures with different uncertainties $S_i$ (Hogg, Bovy, and Lang 2010), so that the logarithm of the likelihood in this case becomes

$$-\ln \mathcal{L} = \frac{n}{2}\ln(2\pi) + \frac{1}{2}\sum_i^n \ln S_i^2 + \frac{1}{2}\sum_i^n \frac{(f(x_i) - y_i)^2}{S_i^2} . \quad (A2)$$

Thus, maximizing the likelihood is equal to minimizing the right-hand side of Eq. (A2). In addition, $\frac{n}{2}\ln(2\pi) + \frac{1}{2}\sum_i^n \ln S_i^2$ is a constant that only depends on the data, not on the model (as a consequence, it is usually omitted when comparing models that depend on the same data, as it cancels out). Then, maximizing the likelihood is equivalent to minimizing $\sum_i^n \frac{(f(x_i)-y_i)^2}{S_i^2}$.

Comparing maximum likelihood of two models is useful if both models have the same number of free parameters. In the other case, we must compare them by the Akaike Information Criterion (Akaike 1974) defined by

$$AIC = 2k - 2\ln \mathcal{L} , \quad (A3)$$

where $k$ is the number of free parameters of the model. The definition of Eq. (A3) supposes a infinitely large number of points to fit by the model, so there is a correction for small size of the sample defined by

$$AIC_C = AIC + \frac{2k^2 + 2k}{n - k - 1} = 2k - 2\ln \mathcal{L} + \frac{2k^2 + 2k}{n - k - 1} . \quad (A4)$$

Thus, given two models fitting the same set of points, the model with lower $AIC_C$ is the best. Nonetheless, models with higher $AIC_C$ should not always be discarded. If we define the difference between the $AIC_C$ of the two models, this is

$$\Delta_i = AIC_{C,i} - AIC_{C,min} , \quad (A5)$$

then, a difference $\Delta_i$ between 0 and 2 indicates that the i-th model has a substantial empirical support (Burnham and Anderson 2002), a difference between 4 and 7 indicates a considerably lesser support of the model, and a difference greater that 10 means that the i-th model has essentially no empirical support and should be completely discarded.

# Appendix B. Figures of extinction threshold depending on the dispersal distance shown as functions of other parameters

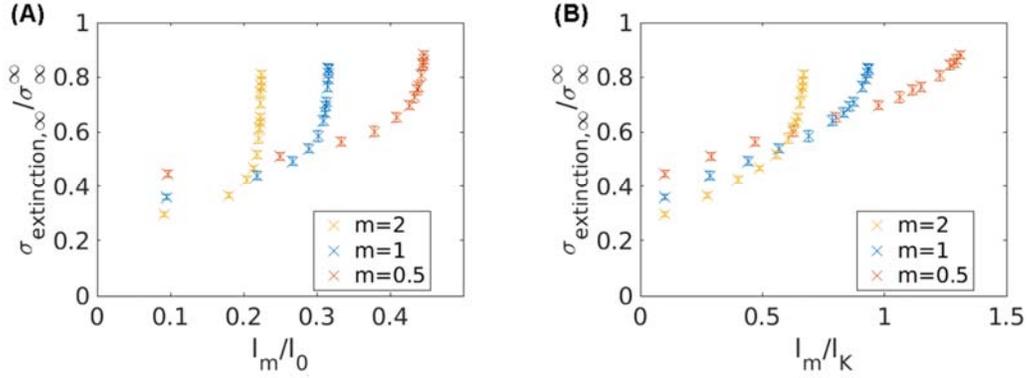

***Figure B1: Extinction thresholds for infinite habitat*** $\sigma_{extinction,\infty}$ *(divided by the respective mean-field value* $\sigma_\infty^\infty$ *for each dispersal rate* $m$*)* ***as a function of the dispersal length*** $l_m$ *divided by the spatial scale of population synchrony around extinction* $l_0$ *(Panel A), and by the spatial scales of population synchrony around carrying capacity* $l_K$ *(Panel B), for different values of the dispersal rate* $m$*. Vertical bars indicate uncertainty in the simulation results. All simulations are for Allee threshold* $A = 0.1$*, carrying capacity* $K = 1$*, extinction rate* $r = 0.1$*, and spatial scale of synchrony of environmental fluctuations* $l_e = 1$*. This figure does not present an approximate scaling behavior, in contrast with the scaling found for the ratio of the dispersal length* $l_m$ *over the spatial scale of synchrony of environmental fluctuations* $l_e$ *(see Fig. 3).*

# Appendix C. Extinction threshold and extinction time.

In all the paper we have estimated the extinction threshold $\sigma_{extinction}$ for a maximum simulation time, or final time for the simulation, of $t = 100 r^{-1}$. At this time we considered the population is extinct if all the positions are below the Allee Threshold. To determine the extinction threshold we plotted the fraction of the positions with a population density above the Allee Threshold $f_A$ as function of the amplitude of environmental fluctuations $\sigma$. The extinction threshold $\sigma_{extinction}$ and its uncertainty was estimated as the center and the width of the range of values where $f_A$ makes the transition between extinct and non-extinct population. (See Panels A and B of Fig. C1 and Appendix D for further details)

Here, we show that extinction threshold mildly decreases as we increase the maximum simulation time, because the extinction time increases as the amplitude of environmental fluctuations decreases. Panel C and D show that the extinction threshold has a relative decrease of the order of 10% when the maximum simulation time is increased by a factor 10.

In the computations of this article, we have considered 10 different realizations for each value of the amplitude of environmental fluctuations $\sigma$. (See, for example, Fig. C1.) This number of realizations is enough for the required precision. For 5 and 10 realizations, the difference in the results for the mean of $\log_{10} t_{extinction}$ for 5 and 10 realizations is small enough (4.8% for small habitat size and 2.9% for large habitat), and even smaller for $\sigma_{extinction}$ (1% difference). In both cases, the differences are smaller than the uncertainties (See Appendix D). Thus, a greater number of realizations would mean only a small difference in the final results.

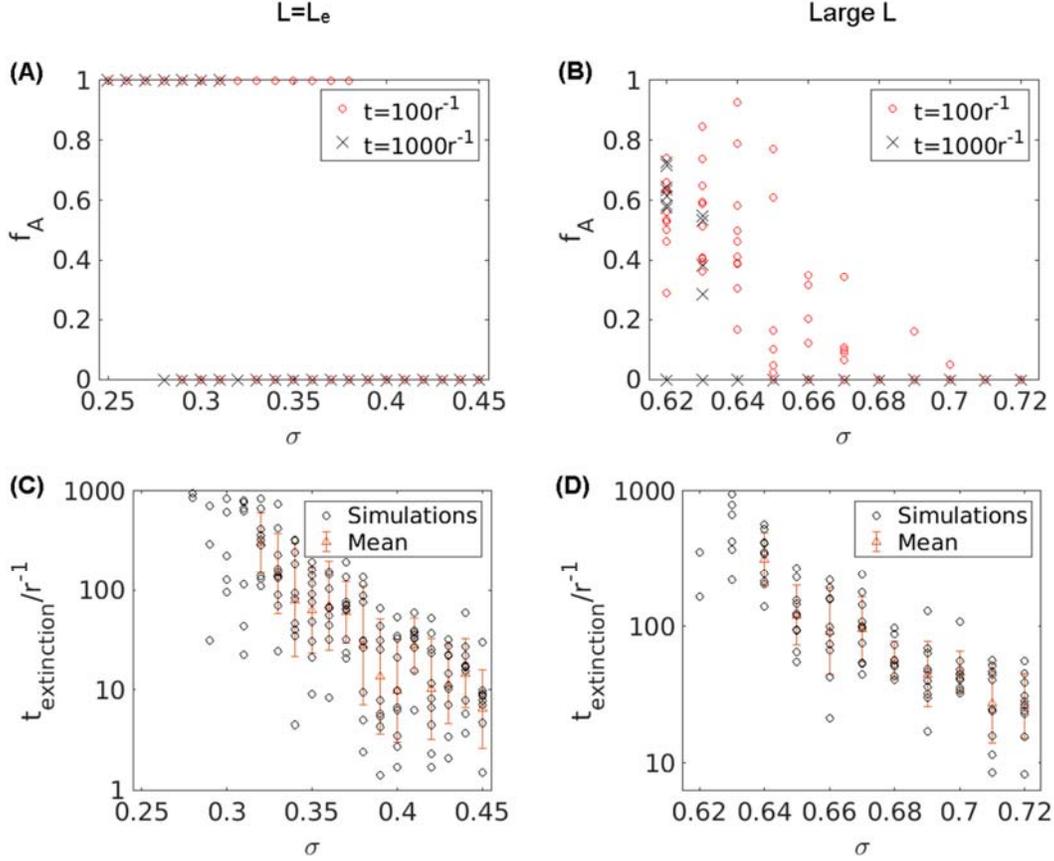

*Figure C1: Extinction threshold for the amplitude of environmental fluctuations σ and extinction times $t_{extinction}$.* Upper panels show the fraction of the positions with a population density above the Allee Threshold $f_A$ for a small habitat size L=$l_e$ (Panel A) and a large habitat size ($L = 187 l_e$) (Panel B) as a function of the amplitude of the environmental fluctuations σ. Bottom panels show the extinction time (in units of the inverse of the extinction rate $r^{-1}$) for a small habitat size (Panel C) and a large habitat size (Panel D) as a function of the amplitude of the environmental fluctuations σ. (Triangles indicate the mean and error bars the 1-standard deviation interval in $\log_{10} t_{extinction}$.) (10 simulations for each value of σ.) For all panels Allee Threshold A=0.1, carrying capacity K=1, extinction rate $r = 0.1$, dispersal rate $m = 1$, and dispersal length $l_m = 0.5\ l_e$ (with the spatial scale of environmental synchrony $l_e$ chosen as the unit of length).

## Appendix D. Calculation of extinction thresholds.

In order to obtain the extinction threshold and its uncertainty for given fixed values of the rest of parameters (i.e., for each point in Figs. 2-6), we have implemented the following procedure. We chose a set of amplitudes of the environmental fluctuations $\{\sigma_1, \sigma_2, ..., \sigma_S\}$, with $\sigma_{i+1} = \sigma_i + \Delta\sigma$, using $\Delta\sigma = 0.01$, and covering a sufficiently large range (such as realizations with the smaller σ never end in global extinction and those with the larger σ always do). Then, we perform $R$ realizations for each $\sigma_i$, and we store the fraction of realizations $k_i$ which finish in global extinction at the end of the simulation (i.e., those realizations with a fraction of the positions with a population density above the Allee Threshold $f_A = 0$ at $t = 100r^{-1}$). Next, we define the discrete distribution

$$p(\sigma_i) = \frac{\min(k_i, 1 - k_i)}{\sum_1^S \min(k_j, 1 - k_j)}, \quad (D1)$$

which satisfies $\sum_1^S p(\sigma_i) = 1$. This distribution can be used to calculate the mean extinction threshold

$$\sigma_{extinction} = \sum_1^S p(\sigma_i) \cdot \sigma_i, \quad (D2)$$

and its standard deviation

$$sd_{biased} = \sqrt{\sum_1^S p(\sigma_i) \cdot (\sigma_i - \sigma_{extinction})^2}. \quad (D3)$$

The distribution is discrete and obtained by a finite sample, so the standard deviation is biased. Only $h \cdot R$ simulations contribute to sampling the distribution $p(\sigma)$, where $h$ is the number of $\sigma_i$ with non-zero $p(\sigma_i)$ (i.e., with $k_i$ different to 0 or 1). The unbiased standard deviation is then given by

$$sd_{unbiased}^2 = \frac{h \cdot R}{h \cdot R - 1} sd_{biased}^2. \quad (D4)$$

Furthermore, the discretization of the $\sigma$ interval in intervals of $\Delta\sigma = 0.01$ can contribute to an underestimation of the standard deviation. To correct this understimation, we computed the uncertainty $S$ of the extinction threshold as

$$S = \sqrt{sd_{unbiased}^2 + \Delta\sigma^2}. \quad (D5)$$

We verified that the interval value of $\Delta\sigma = 0.01$ did not contribute significantly to the uncertainty of the extinction threshold. Thus, the uncertainty $S$ accurately represented the uncertainty in $\sigma_{extinction}$ due to the stochastic nature of the dynamics.

We first considered five realizations, $R = 5$, for each amplitude $\sigma_i$, then we compared the results with those obtained with ten realization, $R = 10$. The average difference between $\sigma_{extinction,R=5}$ and $\sigma_{extinction,R=10}$ was less than 1% and always smaller than the uncertainty $S$. This result indicates that ten realization for each environmental fluctuations amplitude $\sigma_i$ (for each fixed group of value of the other parameters) are enough to accurately calculate the extinction threshold and its uncertainty in the range of parameters that we studied.